\documentclass[aps,prc,groupedaddress,showpacs,twocolumn,floatfix]{revtex4}

\usepackage{graphicx}
\usepackage[dvips]{color}
\usepackage{colordvi}

\def\gtorder{\mathrel{\raise.3ex\hbox{$>$}\mkern-14mu
 \lower0.6ex\hbox{$\sim$}}}
\def\ltorder{\mathrel{\raise.3ex\hbox{$<$}\mkern-14mu
 \lower0.6ex\hbox{$\sim$}}}

\def\beq{\begin{equation}}
\def\eeq{\end{equation}}
\def\ba{\begin{eqnarray*}}
\def\ea{\end{eqnarray*}}

\newcommand{\et}{{\em et al.}}

\begin{document}

\title{Proton $\mathbf{rms}$-radii and electron scattering}

\author{Ingo Sick and Dirk Trautmann}
\affiliation{Dept.~f\"{u}r Physik, Universit\"{a}t Basel,
CH4056 Basel, Switzerland}

\date{\today}
\vspace*{5mm}

\begin{abstract}
The standard procedure of extracting the proton root-mean-square radii from
models for  the Sachs form factors $G_e (q)$ and $G_m (q)$ fitted to elastic 
electron-proton scattering data 
%has a serious flaw. 
is more uncertain than traditionally assumed. The 
 extrapolation of $G(q)$, from the region $q_{min} < q < q_{max}$  
covered by data to momentum transfer $q=0$ where the $rms$-radius is 
obtained, often depends on
uncontrolled properties of the parameterization used. Only when
ensuring that the corresponding densities have a physical behavior at large
radii $r$ can reliable $rms$-radii be determined.  

\end{abstract}


\email{ingo.sick@unibas.ch, Dirk.Trautmann@unibas.ch}
\maketitle

\noindent {\em Introduction. ~~}
Accurate knowledge of the proton root-mean-square ({\em rms}) radii
 is important for both an understanding of proton structure and  the
interpretation of the extremely precise data on transition energies
 in the Hydrogen
atom. Traditionally, the $rms$-radii have been derived from data on elastic
electron-proton scattering at low momentum transfer $q$. The best determinations
are based on  parameterizations of the Sachs form factors $G_e (q)$ and $G_m
(q)$ with the parameters directly fitted to the observables, {\em i.e.} cross
sections and polarization transfer data. The slope of the parameterized 
form factors at $q$=0 yields the charge- and magnetic $rms$-radii, respectively.

This topic of the proton radii has recently become a subject of intense
discussion with the publication of the charge $rms$-radius determined from the 
Lamb shift in muonic
Hydrogen \cite{Pohl10a}. This radius, {0.8409$\pm$0.0004fm}, disagrees by
many standard deviations with the value 0.8775$\pm$0.005fm  from the {\em world} 
data on electron scattering \cite{Mohr10}, 
and it also disagrees with the recent value extracted from the 
 transition energies in electronic Hydrogen \cite{Parthey11}, 
0.8758$\pm$0.0077 fm. This
disagreement has led to a large number of studies dealing with  problems with 
the determination of the radii, or new physics 
\cite{Blunden05}-\nocite{Blunden05b,DeRujula11,Cloet11,Jentschura11b,Jentschura11c,Jentschura11a,Carroll11a}\cite{Carlson11}.

The discrepancy has also led to renewed scrutiny of the procedure used 
to extract 
$rms$-radii from electron scattering data. Problematic  in particular is the
fact that the (e,e)-data sensitive to proton finite size are the ones 
in the region of
momentum transfer $q=$ 0.6$\div$1.2fm$^{-1}$ \cite{Sick07b}; the 
determination of the $rms$-radius
involves an (implicit or explicit) extrapolation to $q=0$.\\[5mm] 
{\em Difficulties due to large-r tail. ~~}
The extrapolation from
finite $q$ to $q=0$ is much more difficult for the proton than for nuclei with
mass number $A>2$. The proton charge and magnetization distributions are  of 
roughly  exponential shape, as the
form factors $G_e(q)$ and $G_m(q)$ are roughly described by  dipole 
distributions. The
exponential tail  at large radii $r$ leads to a very slow convergence of the
proton $rms$-radius as a function of the upper cut-off $r_{cut}$  when
calculated from the  integral
over the charge density $\rho (r)$.  This is
demonstrated in Fig.~1, which compares the convergence to the one obtained for a
heavy nucleus. In order to get 98\% of the proton radius, one has to integrate
out to an  $r_{cut}$ of 3 times the $rms$-radius. 
The uncertainty of this large-$r$ tail, which is poorly fixed
by the (e,e) data, corresponds to uncertainties in the shape of  $G(q)$ at very
 low $q$; this in turn 
\begin{figure}[bht]
\begin{center}
\includegraphics[scale=0.50,clip]{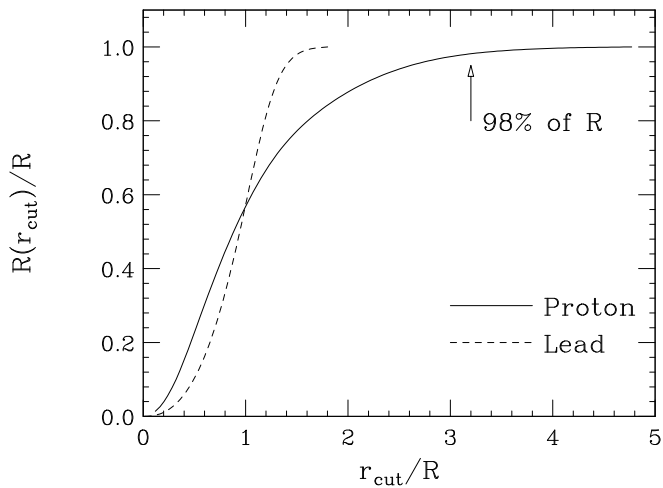}
\parbox{8cm}{\caption[]{$[\int_0^{r_{cut}} \rho(r)~ r^4 dr/ \int_0^\infty \rho(r)~ r^4 dr ]^{1/2}$ as a
function of the upper integration limit  $r_{cut}$.  Both axes are normalized to
the $rms$-radius $R$.
}} 
\end{center} 
\end{figure}
complicates the standard extrapolation of $G(q)$ from the momentum transfers covered by
data to $q=0$ where the radius is determined from the $q=0$ slope.

We illustrate  the difficulties of this extrapolation with several recent
results of form-factor fits. For this {\em qualitative} discussion --- carried
out to understand what happens rather than to derive numerical results --- we
ignore complications such as relativistic effects or two-photon exchange and 
take the  density (charge or magnetic) $\rho (r)$ as  the Fourier transform of 
the corresponding Sachs form factor $G(q)$, and 
vice versa.

As a first example, we consider fits to the   (e,e) data recently measured by
Bernauer \et ~\cite{Bernauer10a,Bernauer10b}  covering the region between 
$q_{min}$= 0.55fm$^{-1}$
and $q_{max}$ = 5.1fm$^{-1}$. These authors used different parameterizations
of $G_e(q)$ and $G_m(q)$ to fit the data and extract the $q=0$ slopes. They showed in
particular the results for fits using an Inverse Polynomial (IP)
parameterization and different orders $N$ of the polynomial. A curious result was
found for the magnetic $rms$-radius, which jumped between order \mbox{$N=7$} 
and $N=10$
from 0.76fm to 0.96fm \cite{Bernauer10a}. The strange behavior of the $N=10$ 
result (which,
incidentally, has the best $\chi^2$ per degree of freedom) turns out to be due to
the fact that at $q>q_{max}$  the $N=10$ IP form factor has a pole. The density
corresponding to such a $G(q)$ has an (oscillatory) tail that extends to
extremely large radii ($r=\infty$). This large-$r$ density implies structure of
 $G(q)$ at very low
$q$, below $q=q_{min}$, and the curvature at very small $q$  falsifies the 
extrapolation to $q=0$ where the radius is extracted.

We must point out that the situation for the $rms$-radius from the  IP 
fit of order $N=7$ (selected by \cite{Bernauer10b}) is not much better; $G_m$ 
has a pole as well, although at
larger momentum transfer with consequences for the density and radius 
that are less severe.

As a second example we mention the form factor fits to the same data 
~\cite{Bernauer10b} made by Lorenz \et ~\cite{Lorenz12a} using a Continued 
Fraction (CF) parameterization. It turns out that the 5-parameter CF fit 
{\cite{Lorenz12b} to the full data set also has a 
pole with correspondingly large values of the density at large $r$,  so the 
 value of the $rms$-radius cannot be relied upon either.

As a third example, we discuss a fit we have made while exploring the role of
the cut-off  $q_{max}$ in the determination of the $rms$-radius. For sake of 
easier comparison, we use the same data, but only up to
$q_{max}$=2fm$^{-1}$; this is entirely sufficient as the data below 
2fm$^{-1}$ are the only ones sensitive to  the $rms$-radii \cite{Sick07b}.
  The data
have been fitted with a 4-parameter Pade parameterization for $G_e$ and $G_m$. 
The fit gives
an excellent $\chi^2$ of  1.06 per degree of freedom, as low as a spline fit to
the data \cite{Bernauer10a}. The fitted $G(q)$  has no pole, but yields a 
charge 
$rms$-radius of 1.48 fm! As for the previous examples, the problem is a
consequence of an uncontrolled behavior of the form factor for $q>q_{max}$
which, together with $G(q_{min}<q<q_{max})$, implies a shape of $\rho (r)$ that has
 a tail to extremely large $r$. 

This unreasonable result can be understood by looking at Fig.~2
which shows the form factor $G_e (q)$ in the region $q<q_{min}$.  The solid
curve represents the Pade fit,  the dotted curve corresponds to a usual
parameterization yielding an $rms$-radius of $\sim$0.88fm from a correspondingly
smaller $q=0$ slope. One can conceptually split the Pade
curve into two additive contributions $G_1$ and $G_2$. $G_1 (q)$ corresponds 
to the Pade fit for
$q^2>$ 0.06fm$^{-2}$, supplemented for $q^2<0.06$fm$^{-2}$ by the dashed curve; 
 $G_2 (q)$
corresponds to the difference of the Pade form factor and $G_1 (q)$. The form
factor $G_1 (q)$ has a norm of $\sim$0.995 and a ``normal'' slope at $q=0$
corresponding to an $rms$-radius of $\sim$0.88fm. The $G_2 (q)$ term looks, 
roughly
speaking, like a gaussian $e^{-q^2/a^2}$ with $a^2 \sim$ 0.02fm$^{-2}$ and a
norm of $\sim$0.005. In $r$-space, this $G_2$ corresponds to a density of 
Gaussian shape, proportional to  
$e^{-r^2/(200\mbox{\small{fm}}^2)}$.  This term extends to extremely large values of $r$.
Its contribution at large $r$ leads, despite the small overall norm, to the
unreasonable $rms$-radius of 1.48fm.

\begin{figure}[hbt]
\begin{center}
\includegraphics[scale=0.48,clip]{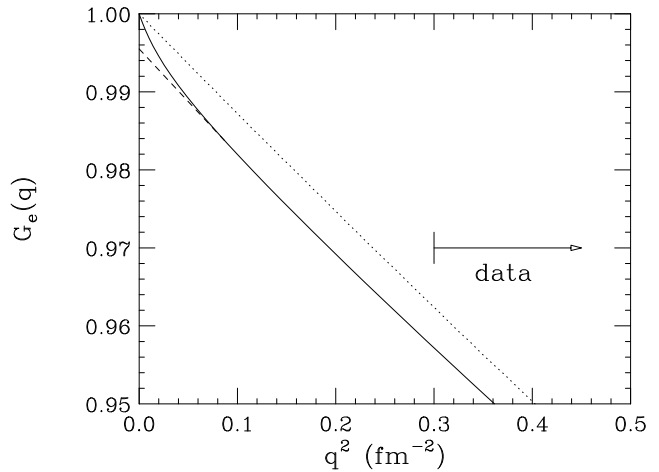}
\parbox{8cm}{\caption[]{Pade fit to the data $q<2$fm$^{-1}$ of
Ref.~\cite{Bernauer10b} (solid curve) yielding an $rms$-radius of 1.48fm.
A ``standard'' fit to the same data, producing an $rms$-radius of
$\sim$0.88fm, is given by the dotted curve. (Note that the normalization of the 
data (not shown) is floating, so both fits
give excellent $\chi^2$ when compared to data normalized to the respective fit).
 The form factor $G_1 (q)$ corresponds to the solid
curve for $q^2 > $ 0.06fm$^{-2}$, supplemented for $q^2 < $ 0.06fm$^{-2}$ by 
the  dashed curve. $G_2 (q)$
corresponds to the difference between the solid curve and $G_1(q)$.
}\label{}} 
\end{center} 
\end{figure}

From these examples it becomes clear that the usual way of determining the
proton  $rms$-radii by extrapolating the data from $q_{min} < q < q_{max}$ to 
$q=0$ via a
parameterization of $G(q)$ is  unreliable. Even for standard 
parameterizations such as the Pade parameterization, with no obvious faults such as  poles 
 at $q>q_{max}$  or
 a divergence at large $q$ such as present in the popular expansions of $G(q)$
in terms of  powers of  $q^2$,
unreasonable results can be generated. Such  procedures  cannot be trusted to
yield the physical $rms$-radius.

\noindent
{\em  Density at large r}. ~~} 
 In order to better understand the origin of
the difficulty, it is helpful to confront the approaches used for the 
determination of $rms$-radii for
nuclei with mass number $A>2$ and $A\leq 2$. For $A>2$ the density is
parameterized, the observables are calculated via a solution of the Dirac
equation and the parameters are adjusted for good $\chi^2$. The $rms$-radius is
obtained from an integral over $\rho (r)$ or, equivalently, from the $q=0$
slope of the corresponding Born approximation form factor. For $A \leq 2$, on 
the other hand, 
 the form factors $G(q)$ in general are parameterized directly in $q$-space, 
the observables are calculated with/without corrections
beyond first Born approximation and the parameters adjusted for the best
$\chi^2$. The slope of $G(q)$ at $q=0$  gives the $rms$-radius.

These two approaches are {\em not} equivalent. When parameterizing $\rho (r)$
using {\em e.g.} Fermi densities, Sum-Of-Gaussians (SOG) or Fourier-Bessel (FB)
series an implicit assumption is made: that $\rho (r) = 0$ outside some
radius $r_0$, or that $\rho (r)$ falls off  exponentially at large
$r$. This ``assumption'' reflects our physics understanding of nucleon and
nuclear wave functions: the density  at large enough $r$ must approach
 zero. 
%Ignoring for the moment complications mentioned below, we expect the large-r
%density to fall off like a  Whittaker function $W^2 (\kappa r)/r^2$, with 
%$\kappa$ being given by the removal energy of the least-bound charged
%constituent.

When parameterizing $G(q)$ directly in $q$-space, this condition on $\rho (r)$ 
at large $r$ is {\em not} used. Depending on the parameterization chosen for
$G(q)$ totally unreasonable behaviors of $G(q>q_{max})$ and the corresponding
$\rho (r)$ at large $r$ may occur,  and these are responsible for the erroneous
$rms$-radii. This is clearly a serious  deficiency  that calls for a different
approach. 

The problem induced by an uncontrolled behavior of $G(q>q_{max})$ is less
severe for fits that use the data
up to the largest momentum transfers  where cross sections and polarization transfer data 
are available. In 
these cases, the values of  the fitted $G(q>q_{max})$ are  strongly  
constrained by the small values of  the  $G(q)$'s near the top end 
of the $q$-range covered by experiment. Parameterizations that enforce in 
addition a fall-off of at least  $1/q^4$  as required to get a regular density
at the origin, then help to ensure 
that  the values of $G(q>q_{max})$ remain small and that the data constrain the
shape of $\rho (r)$ (including its tail) as much as possible.

From the parameterization of $G(q)$ alone it is very difficult to judge the
behavior  $\rho (r)$ other than by looking simultaneously at
the corresponding density (provided the parameterization of $G(q)$ 
does have a Fourier
transform!). As a matter of fact, most published fits to e-p data have not been
checked for the large-$r$ behavior of $\rho$, and   often yield 
unreasonable behavior as the above examples show. 

Of the parameterizations that have been used in the past to fit the e-p
scattering data, there are basically only two types that  constrain the 
large-$r$ behavior of the density: On the one hand side there are the fits based 
on  the Vector Dominance Model (VDM) supplemented by 2$\pi$ exchange
\cite{Lorenz12a}\nocite{Lorenz12b,Iachello73,Hoehler76,Mergel96,Lomon01}-\cite{Hammer04}. 
They implicitly 
constrain the large-$r$ fall-off through the masses of the  
exchange particles that are assumed to mediate the photon-nucleon 
interaction. These VDM-type fits, however,  have in general produced
 values of $\chi^2$ that are significantly larger than those achieved by 
more phenomenological parameterizations and they show systematic deviations from 
the data at low momentum transfer; the model seems to lack the freedom
required for a good fit of the e-p data. The radii extracted from  a poor fit
obviously cannot be trusted. 

The other approach that ensures a good behavior of  the large-$r$ 
density is the one involving fits with  SOG densities fitted to both the e-p
data and a calculated shape of the  large-$r$  density \cite{Sick12}.

To remedy the unsatisfactory situation with extrapolation of $G(q)$ to $q=0$
one can proceed as described below. This procedure, unfortunately, 
 is more involved than a simple fit of (e,e)-data using a convenient
parameterization for $G(q)$, but it helps  to avoid the pitfalls discussed 
above.
   
   In order to guarantee a sensible behavior of the density at large radii, one
should employ parameterizations that have an easily accessible form  
%density in $r$-space.  When choosing 
%parameterizations that have analytical Fourier transforms, one can fit the 
%parameters directly in $q$-space without taking too literally the form factor 
%as a Fourier transform of a density. 
 in both $r$- and $q$-space. One then can easily check the r-space 
 behavior while fitting the parameters directly in $q$-space without taking 
 too literally the form factor as a Fourier transform of the density. 
  
 For the proton (and the same is true for the deuteron, see \cite{Sick01})  the
pronounced sensitivity of the $rms$-radius to the tail of the density can be
reduced by using a physical model for the large-$r$ density. At large distances,
the density of  any bound multi-constituent system is dominated  by the
least-bound Fock  component, in the case of the proton the $\pi^+$ from the 
neutron + pion ($n+\pi^+$)  configuration. The corresponding (relative) density
can be easily calculated; sophistications  such as inclusion of relativistic
effects or two-photon exchange (which affect the relation between density and
form factor), pion finite size, $\Delta + \pi$ components {\em etc} can be
incorporated  (their numerical effect on the tail-shape has been found to be
relatively  small \cite{Sick12}). The resulting density can be used in the fit
to constrain the shape of  $\rho (r)$ at very large $r$ where the density is
safely in the asymptotic regime. In this approach the most dangerous aspect of
the  model-dependence  ---  implicit assumptions on the large-$r$ behavior due
to the choice of the model density or  model form factor which affect the
curvature of $G(q)$ below $q_{min}$  ---  is replaced by  a tail constraint that
is explicitly stated and can be included with appropriately specified
uncertainties. 

When using this type of  approach  one can employ rather general multi-parameter 
expansions of the density/form factor such as Hermite and Laguerre
polynomials times their weight function, or SOG densities; these  allow to fit 
with good $\chi^2$  the data over 
the  full $q$-range where data are available. As we have discussed above, this 
is most desirable in order
to also constrain  as much as possible the shape of the density, including the 
large-$r$ tail,  by the e-p data themselves.     

 A fit of the {\em world} data following the above procedure has been carried 
out in \cite{Sick12} and yielded an $rms$-radius of  0.886$\pm$0.008fm.

{\em Conclusions. ~~}
While the standard parameterizations of the nucleon form factors $G_e (q)$ and
$G_m (q)$ fitted to e-p scattering data are valid representations of the data
in the $q$-region where they have been measured,
they are not suitable for an extrapolation to $q=0$ where the proton $rms$-radii
are extracted. These parameterizations of the $G(q)$'s lack the constraint that
the corresponding 
densities must approach zero at large $r$ in a way that is compatible with
our physics understanding of the proton. Most published fits do not respect this
constraint and often produce unreliable radii. 
  
%With the choice of a $q$-space parameterization for which then density  has a 
%well-controlled behavior at large $r$ this condition can be imposed, and 
%much more reliable radii be determined. 

Constraining the large-radius tail of the density using physical arguments is
bound to yield a more reliable value for the physical $rms$ radius.

%\bibliographystyle{unsrt}
%\bibliography{/usr/users/sick/sum3}

\begin{thebibliography}{10}
\bibitem{Pohl10a}
R.~Pohl, A.~Antognini, F.~Nez, F.D. Amaro, F.~Biraben, J.M.R. Cardoso, D.A.
  Covita, A.~Dax, S.~Dhawan, L.M.P. Fernandes, A.~Giesen, T.~Rraf, T.W.
  H{\"a}nsch, P.~Indelicato, L.~Julien, C-Y. Kao, P.~Knowles, J.A.M.Lopes,
  E-O.~Le Bigot, Y-W. Liu, L.~Ludhova, C.M.B. Monteiro, F.~Mulhauser, T.~Nebel,
  P.~Rabinowitz, J.M.F dos Santos, L.~Schaller, K.~Schuhmann, C.~Schwob,
  T.~Taqqu, J.F.C.A. Veloso, and F.~Kottmann.
\newblock {\em Nature}, 466:213, 2010.

\bibitem{Mohr10}
P.~Mohr, B.~Taylor, and D.B. Newell.
\newblock {\em Rev. Mod. Phys.}, 84:1527, 2012.

\bibitem{Parthey11}
C.G. Parthey, A.~Matveev, J.~Alnis, B.~Bernhardt, A.~Beyer, R.~Holzwarth,
  A.~Maistrou, R.~Pohl, K.~Predehl, T.~Udem, T.~Wilken, N.~Kolachevsky,
  M.~Abgrall, D.~Rovera, C.~Salomon, P.~Laurent, and T.W. H\"ansch.
\newblock {\em Phys. Rev. Lett.}, 107:203001, 2011.

\bibitem{Blunden05}
P.G. Blunden, W.~Melnitchouk, and J.A. Tjon.
\newblock {\em Phys. Rev. C}, 72:034612, 2005.

\bibitem{Blunden05b}
P.G. Blunden and I.~Sick.
\newblock {\em Phys. Rev. C}, 72:57601, 2005.

\bibitem{DeRujula11}
A.~deRujula.
\newblock {\em Phys. Lett. B}, 697:26, 2011.

\bibitem{Cloet11}
I.C. Clo{\"e}t and G.A. Miller.
\newblock {\em Phys. Rev. C}, 83:012201, 2011.

\bibitem{Jentschura11b}
U.D. Jentschura.
\newblock {\em Ann. Phys.}, 326:516, 2011.

\bibitem{Jentschura11c}
U.D. Jentschura.
\newblock {\em Ann. Phys.}, 326:500, 2011.

\bibitem{Jentschura11a}
U.D. Jentschura.
\newblock {\em arXiv:1107:1737}, 2011.

\bibitem{Carroll11a}
J.D. Carroll, A.W. Thomas, J.~Rafelski, and G.A. Miller.
\newblock {\em Phys. Rev. A}, 84:012506, 2011.

\bibitem{Carlson11}
C.E. Carlson and M.~Vanderhaeghen.
\newblock {\em arXiv:1109.3779}, 2011.

\bibitem{Sick07b}
I.~Sick.
\newblock {\em Can. J. Phys.}, 85:409, 2007.

\bibitem{Bernauer10a}
J.C. Bernauer.
\newblock {\em Thesis, Univ. of Mainz}, 2010.

\bibitem{Bernauer10b}
J.C. Bernauer, P~Achenbach, C.~Ayerbe Gayoso, R.~B{\"o}hm, D.~Bosnar,
  L.~Debenjak, M.O. Distler, L.~Doria, A.~Esser, H.~Fonvieille, J.M. Friedrich,
  M.~G{\,o}mez~Rodr{\,i}guez de~la Paz, M.~Makek, H.~Merkel, D.G. Middleton,
  U.~M{\"u}ller, L.~Nungesser, J.~Pochodzalla, M~Potokar, S.~{S{\'a}nchez
  Majos}, B.S. Schlimme, S.~\v{S}irca, Th. Walcher, and M.~Weinriefer.
\newblock {\em Phys. Rev. Lett.}, 105:242001, 2010.

\bibitem{Lorenz12a}
I.T. Lorenz, H.-W. Hammer, and U.-G. Meissner.
\newblock {\em Eur. Phys. J. A}, 48 11:151, 2012.

\bibitem{Lorenz12b}
I.T. Lorenz.
\newblock {\em priv. com.}, 2012.

\bibitem{Iachello73}
F.~Iachello, A.D. Jackson, and A.~Lande.
\newblock {\em Phys. Lett.}, 43B:191, 1973.

\bibitem{Hoehler76}
G.~Hoehler, E.~Pietarinen, I.~Sabba-Stefanescu, F.~Borkowski, G.G. Simon, V.H.
  Walther, and R.D. Wendling.
\newblock {\em Nucl. Phys. B}, 114:505, 1976.

\bibitem{Mergel96}
P.~Mergell, U.-G. Meissner, and D.~Drechsel.
\newblock {\em Nucl. Phys. A}, 596:367, 1996.

\bibitem{Lomon01}
E.L. Lomon.
\newblock {\em Phys. Rev. C}, 64:035204, 2001.

\bibitem{Hammer04}
H.-W. Hammer, D.~Drechsel, and U.-G. Meissner.
\newblock {\em Phys. Lett. B}, 586:291, 2004.

\bibitem{Sick12}
I.~Sick.
\newblock {\em Prog. Part. Nucl. Phys.}, 67:473, 2012.

\bibitem{Sick01}
I.~Sick.
\newblock {\em Prog. Nucl. Part. Phys.}, 47:245--318, 2001.


\end{thebibliography}

\end{document}